%% file: STQN-CM-V1.tex
\documentclass[12pt, final, twocolumn, twoside, romanappendices]{IEEEtran}

\usepackage{amsmath,amssymb}

\usepackage[bookmarks,colorlinks]{hyperref} 
\hypersetup{colorlinks,citecolor= red,filecolor= blue,linkcolor= blue,urlcolor=blue}

\usepackage{graphicx}%
\usepackage[usenames,dvipsnames]{xcolor}

\usepackage{tikz,pgfplots}

\usepackage[printonlyused,withpage]{acronym}
\usepackage{cite} 
\usepackage{color}
\usepackage{comment} 

\usepackage{WINS-LatexInclusion/Preambles/winsnotation} 
\usepackage[level=3]{WINS-LatexInclusion/Preambles/wgroup_message}  
\input{WINS-LatexInclusion/Acronyms/WINS-Acronyms.tex}

\input{WINS-LatexInclusion/Definitions/WINS-Format-Def.tex}
\input{WINS-LatexInclusion/Definitions/WINS-Symbol-Def.tex}
\input{WINS-LatexInclusion/Definitions/WINS-Format-Def} 
\input{WINS-LatexInclusion/Acronyms/WINS-Acronyms} 

\usepackage[yyyymmdd,hhmmss]{datetime} 
\newdateformat{monthyeardate}{\monthname[\THEMONTH] \THEDAY, \THEYEAR} 
\usepackage[pagewise,mathlines,displaymath]{lineno} 

\newcommand{\paperTitle}{Satellite-Terrestrial Quantum Networks and the Global Quantum Internet}

\pgfplotsset{compat=1.18}
\begin{document}

\twocolumn



\title{\paperTitle}


\author{
	\vspace{0.2cm}
	
   Andrea Conti,
Robert Malaney, and
   Moe~Z.~Win\\
    \thanks{
    }    
    
    \thanks{
     Andrea Conti is with the Department of
Engineering and CNIT, University of Ferrara, 44122 Ferrara, Italy.

        Robert Malaney (corresponding author) is with the School of Electrical Engineering and Telecommunications, University of New South Wales, Sydney, Australia.
    
        Moe~Z.~Win is with
        the Laboratory for Information and Decision Systems, 
        Massachusetts Institute of Technology, 
         Cambridge, MA 02139,
        USA.
   
       \noindent \textit{Published version: \href{https://doi.org/10.1109/MCOM.007.2300854}{https://doi.org/10.1109/MCOM.007.2300854}}.
	}		
}

\maketitle 
\begin{abstract}
This paper will explore the design and implementation of quantum networks in space integrated with quantum networks on Earth.
We propose a three-layer approach, involving GEO and LEO satellites integrated with terrestrial ground stations. We first analyze the  channel conditions  between the three layers, and then highlight the key role of LEO satellites in the integrated space-terrestrial system - namely the source of entanglement distribution
between specified terrestrial stations via direct downlink quantum-optical channels. The GEO satellites in the considered system are used primarily as `coordination' stations, managing and directing the LEO satellites regarding the positioning and timing of entanglement distribution. 
Complexity, in the form of entanglement distillation and quantum-state correction, is concentrated at the terrestrial stations, and teleportation is used as the primary quantum channel in the LEO uplinks and the inter-terrestrial channels. 
Although our designs are futuristic in that they assume limited quantum memory at the transceivers, we also discuss some near-term uses of our network in which no quantum memory is available. 
\end{abstract}

\begin{IEEEkeywords}
Quantum networks, quantum state, quantum channel, entanglement distribution, next-generation networks.
\end{IEEEkeywords}

\acresetall		


\section{Introduction}\label{sec:intro}
\IEEEPARstart{Q}{uantum networks}  are expected to be  ubiquitous, and complimentary to classical networks, in the next next decade, with many large-scale terrestrial quantum networks currently being designed and built. These range from from city-wide to intercountry-wide  networks, such as those being constructed in Europe~\cite{euro}. Beyond this, satellite-based quantum networks are also being considered~\cite{ref2},  boosted by the outstanding success of the Micius satellite that removed any doubt that space-based quantum communication with Earth is viable~\cite{ref3}. The use of satellites as a backbone to a global quantum communication network is motivated to a large extent by the favourable loss conditions in  free space with respect to losses in optical fibers for distances beyond 100\,km.

In spite of the above advancements in quantum communication deployments, the design principles underpinning the global quantum Internet (GQI) are still to be fully agreed upon. The GQI aims to be a multi-functional complex next-generation network, accommodating a wide variety of protocols and applications. It will serve as the security guardian of all future communications, likely via a hybrid of Post-Quantum Cryptography solutions interleaved with quantum key distribution (QKD). Such a hybrid encryption scheme will deliver the ultimate security guarantee; state-of-the-art classical encryption enhanced with the information-theoretic security afforded by QKD when conditions allow.

The GQI must also also serve as the basis of the inter-communication between various forms of quantum-enhanced sensors, be the link between all future quantum computers currently under worldwide development, and serve as the basis for quantum applications yet to be discovered. Beyond these demanding attributes, the GQI will have to accommodate multiple information  technologies - due to the ability to conduct quantum information processing in both the discrete variable (DV) domain and the continuous variable (CV) domain. DV and CV technologies present pros and cons depending on specific applications, performance requirements, and channel conditions.
DV have been currently deployed, but we believe that going forward both technologies will be deployed in the GQI, with various mechanisms deployed to integrate them across both space-based and terrestrial based quantum networks. 

This paper investigates the overall design principles of the GQI, paying particular focus on the optimization of such a complex network. DV and CV technologies are covered by our designs. The  design principles we will consider are summarized in the following. We utilize high-quality optical links when available for quantum communication and radio communication for all classical communications;  channel conditions in quantum communication are asymmetric in many situations and this unusual circumstance should be exploited whenever possible; complexity in the quantum domain should be kept on terrestrial situation whenever possible; and hybrid quantum communication should be utilized when circumstances demand it. 

\section{Quantum communications in  satellite-terrestrial networks}\label{sec:STQcom}
\begin{figure}
	\centering
	\includegraphics[width=.95\linewidth]{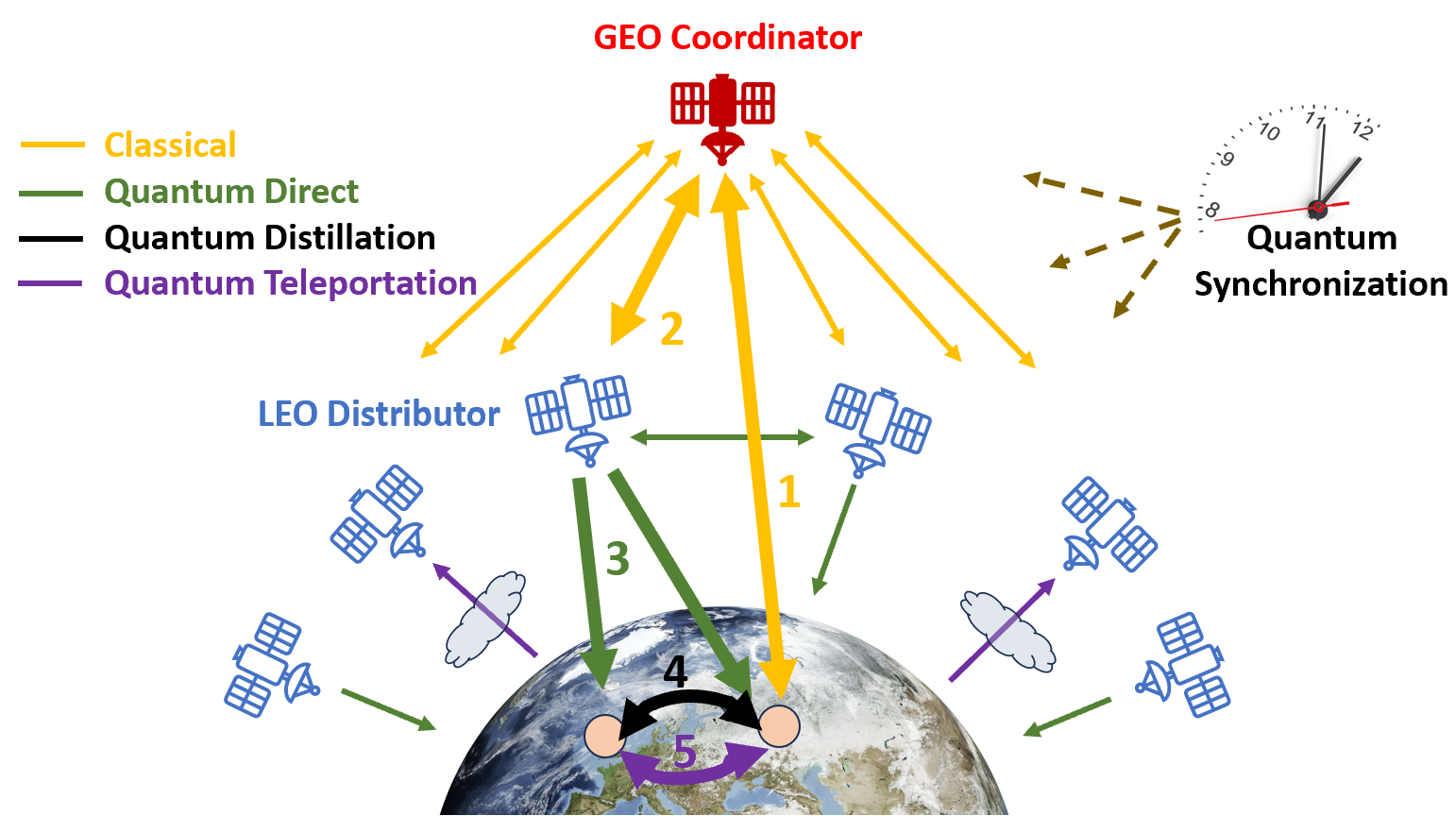}
	\caption{\textbf{A multilayered satellite network for the global quantum Internet.} The GEO satellites are used for coordinating the quantum information transfer between different LEO satellites, as well as between LEO satellites and terrestrial stations. A potential sequence of communication is as follows: (1) A request  is initiated from a terrestrial station (right pink circle) to undertake quantum communication with another terrestrial station (left pink circle). (2) A request is then sent from the GEO satellite to the selected LEO satellite to (3) commence entanglement distribution in two separate downlinks to the two terrestrial stations. (4) A sequence of entanglement distribution then commences between the two terrestrial stations. This distillation process involves quantum measurements at the two stations in conjunction with classical communications between the stations, and leads to a reduced set of entangled qubits between the stations at close-to-maximal entanglement. (5) Teleportation  between the terrestrial stations then ensues, in which qubits can be transferred from one station to the other. Note, in the above the classical communications between satellites can be at optical or radio frequencies, whereas the quantum information is assumed to be at optical frequencies only. The terrestrial stations can classically communicate directly  with each other via radio frequencies. It is also assumed the entire network is highly synchronized, possibly via  quantum-enhanced synchronization techniques. Not shown are other GEO satellites which could form part of the network,}
	\label{fig:network}
\end{figure}
We propose an integrated design for  combined inter-satellite satellite-terrestrial networks. In doing this, we will consider a finite lifetime for any distributed entanglement and quantum memories at transceivers with storage time greater than the round-trip communication time.
That is, a quantum memory is available at all transceivers with a coherence (storage) time at least of order 1 second (less than the  round-trip communication time for any communication link in the network). We point out, however, many functions within the network require no quantum memory, \textit{e.g}., entanglement generation.
The design philosophy follows three main principles - (i) quantum communications will always seek to use the channel  with lowest predicted losses, (ii) complex photonic engineering such as entanglement distillation  is deployed only on ground stations, (iii) unless otherwise stated, all non-quantum communications are radio-based.
In  following the first of these principals,  in very lossy channels we will make use of teleportation instead of direct quantum information transfer.We stress, that our proposed architecture is but one of several possible architectures \cite{costs} - most of which involve only LEO satellites and ground stations. In this article, we argue for the benefits brought by use of  geosynchronous Earth orbit (GEO) satellites within the overall architecture.

\begin{figure}
	\centering
	\includegraphics[width=.95\linewidth]{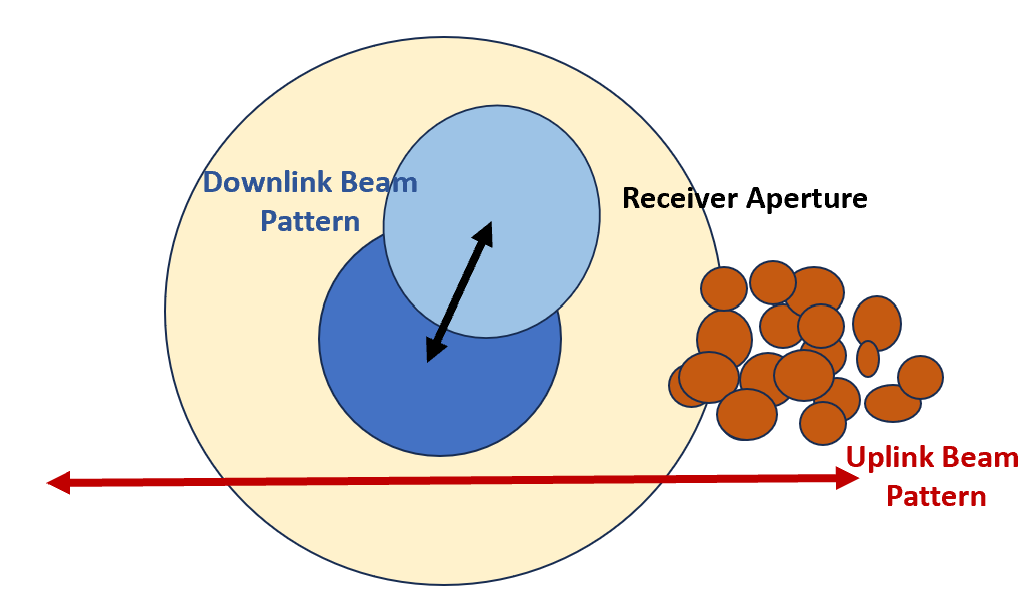}
	\caption{\textbf{Uplink-Downlink Beam Patterns.} Shown here are the anticipated beam patterns and movements for the different directions through the Earth's atmosphere to-and-from LEO. The downlink is anticipated to show much less movement and deformation relative to the uplink. In the latter link, beam wandering and other distorting effects are likely to lead to a speckle patter than move in and out of the receiving aperture. In the downlink these issues are anticipated to be much less pronounced. The major reason for this asymmetry  can be traced back to the ratio of beam-size to turbulent-cell-size at entry to the atmosphere - in the downlink this ratio is much larger relative to the uplink. It is for this reason we believe that in the uplinks to LEO from terrestrial stations, teleportation  will be the preferred mode of quantum communication -  the entanglement distribution needed for the teleportation being supplied by the downlink. The patterns shown in this figure are consistent with the predictions from many theoretical studies, as well as detailed phase screen simulations. }
	\label{fig:links}
\end{figure}

A schematic of our integrated space-terrestrial quantum network is shown in Fig.~\ref{fig:network}.
As can be seen, GEO  satellites form part of this network. However, as opposed to other proposed schemes that use such high-orbit satellites as part of the quantum signalling, we deem the anticipated losses from GEO  ($60$\,dB for anticipated satellite dimensions) to render quantum communication largely unattainable (see later discussion). In our proposal we use GEO satellites  as `management' nodes that utilize radio-based classical signalling only. The low Earth orbit (LEO) satellites are the key elements of this architecture, instigating the downlink entanglement distribution to the terrestrial ground stations, as well as providing the main space-based quantum communication links. Quantum communication, in the uplink from ground station  to LEO satellite, will utilize teleportation due to the high losses predicted for such channels.  In the terrestrial quantum links, teleportation will be used as the primary means of quantum communication,  with entanglement swapping being used only when  necessary. The architecture considered will assume that no terrestrial flying vehicles form part of the network, and that all quantum communication utilize optical frequencies only.  All classical communication across our integrated network  will be radio-based - unless stated otherwise. Note that an optional channel for the classical communication is optical-based, as encoded classical information in such channels can accommodate large losses yet still be fully functional (e.g., simple on-off keying). 

\subsection{Quantum Channels}
To make further progress in the design of a GQI we first must set typical  optical losses anticipated for our quantum channels
The design of a CGI requires modeling the quantum channels (i.e., optical losses); classical channels are considered protected via classical-channel coding techniques. To determine such losses we must first set what we anticipate to be the aperture diameter of transmitter and receivers (collectively, the transceivers). Here, unless stated otherwise, we will typically  assume terrestrial and satellite diameters to be in the range 0.5-2.5\,m and 0.1-0.4\,m, respectively. We will also take LEO  to be at 500\,km -1200\,km from ground level and inter-satellite separation in LEO to be $200$\,km. Such aperture and distances are widely used in other studies, however, the losses we next describe can be easily adjusted for different aperture-distance combinations. 

\subsubsection{LEO-Terrestrial Channels} 

Distinct from traditional radio communications, wireless optical communications to and from LEO can be severely influenced by atmospheric conditions, most notably air-turbulence (other effects such as absorption can be largely negated by optical wavelength choices). This is particularly so in the relevant quantum domain, and much theoretical work  has been devoted to the development of an understanding on how quantum information is affected by passage through Earth's atmosphere, \textit{e.g.}, ~\cite{ref6}, and based on previous works the following basic expectations can be drawn. The uplink channel (Terrestrial-to-LEO) is heavily affected by beam wander effects, rendering the optical transmitted beam to misalign with the receiver aperture for some time periods. This misalignment can  lead to a partial overlap with the beam on receiver aperture or with no beam overlap at all. The movement of the beam is anticipated to have a coherence time of order one millisecond and leads to a fading behavior for the channel. For anticipated transceiver apertures such beam movement is expected to lead to average losses of order 20\,dB~\cite{ref3}.

A peculiar feature of the optical channel in the LEO-Terrestrial setting is the asymmetrical behavior of the channel. In the downlink (LEO-to-Terrestrial) the channel losses are anticipated to be significantly smaller, around $\sim 5$\,dB. Indeed, for large ground telescopes used as receivers, these losses can be effectively negated  \cite{ref6}. The principal reason for this loss asymmetry in the uplink and downlink channels is the relation of the beam widths to turbulent cell dimensions at atmospheric entry point. In the downlink,  the beam width is typically  larger than turbulent cell dimensions at entry to the atmospheric layer. In the uplink the opposite is true - the beam enters the atmosphere immediately after leaving the transmitter and is not much larger than its small beam waist. The losses in the downlink are therefore hardly affected by beam wandering and largely  set by diffraction losses alone.  This means as opposed to the fading uplink, the downlink can be considered as a fixed channel - an outcome we can exploit in GQI designs. More specifically, this leads us to use the downlink for all entanglement distribution between the terrestrial ground stations, and the use of teleportation for any required uplink quantum communications. The role of turbulence in the uplink and downlink communications are captured in Fig.~\ref{fig:links}.

\begin{figure}
	\centering
	\includegraphics[width=.95\linewidth]{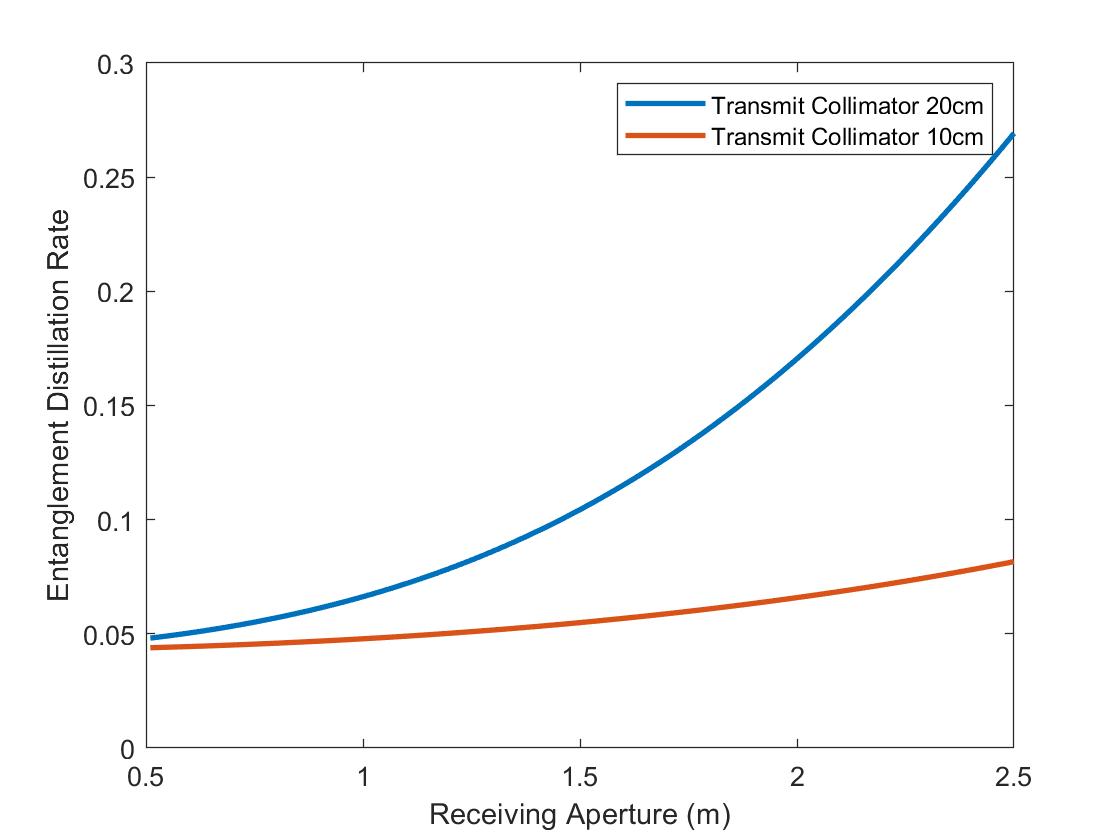}
	\caption{\textbf{LEO  Entanglement Generation.} Shown here is the rate of entanglement distillation - the number of `ebits' generated per channel use - as a function of transceiver apertures (an ebit represents one maximally entangled Bell pair). Here, we have taken the the downlink (1200\,km) loss distribution to be modelled 
 as a function of a
parameter $b$, which for $b$ approaching zero is a fixed loss set by aperture sizes, which has a Gaussian tail behavior
with deviation $b$. In the calculations shown, we have set $b = 0.1$,
 and utilized the reverse coherent information - a metric known to be an achievable rate for the distillation rate \cite{ref5}. Our resulting loss distribution function is consistent with the phase screen simulations for the downlink channel reported in \cite{ref6} - simulations that show the downlink is largely dominated by diffraction losses. Without the distillation process invoked there would be effectively a zero rate of ebit-generation unless losses were largely negated by the use of very large receiving apertures and transmit collimator widths (none of which are represented in this figure). Note, the results shown here are optimal results, derived from information theoretic constructs assuming loss as the only noise process. As such, they should be viewed as upper limits to the ebit-generation possible. }
	\label{fig:rates}
\end{figure}

\begin{figure}
	\centering
	\includegraphics[width=.95\linewidth]{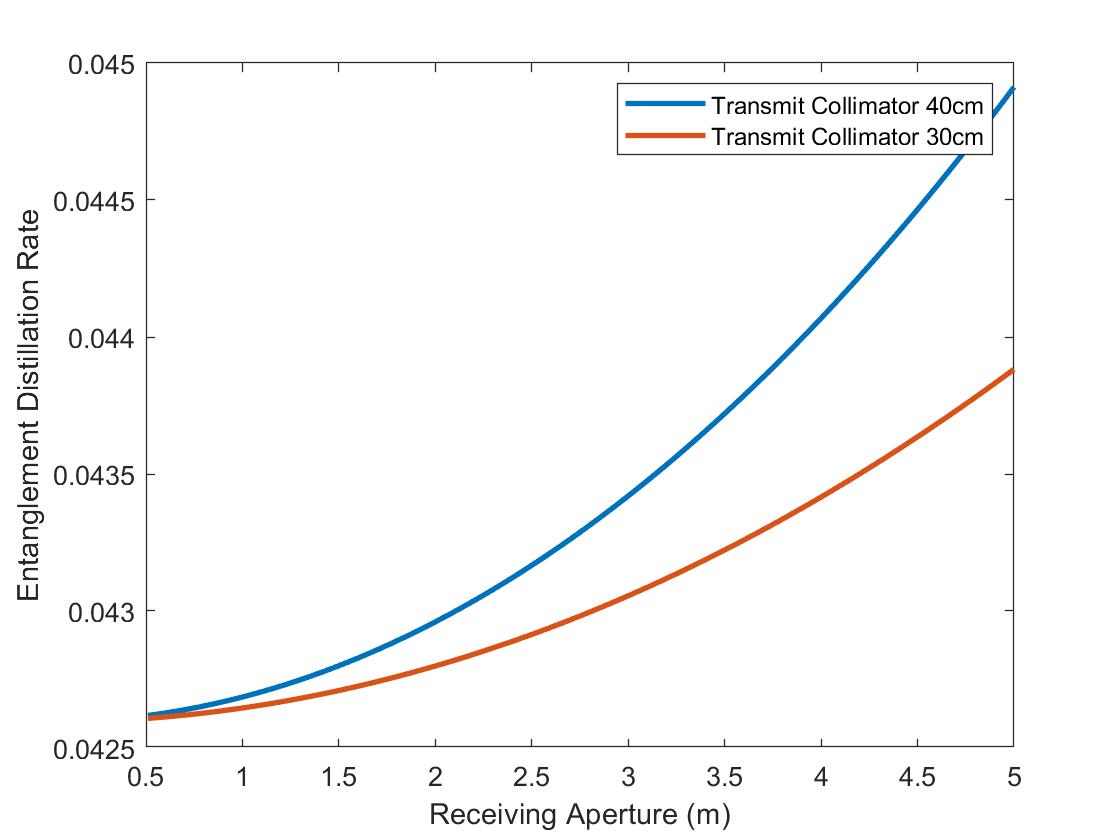}
	\caption{\textbf{GEO  Entanglement Generation.} Here, the distance between the satellites participating in the entanglement distribution is 36000\,km. We can see in spite of large collimator and receiving apertures the entanglement generation remains below 0.05, much smaller than most of the LEO link (1200\,km) parameter space (in these calculations all other parameter settings are the same as the LEO link). It is for this reason we believe the GEO link will primarily be utilized for classical communications within the broader GQI architecture.}
	\label{fig:rate2s}
\end{figure}

\subsubsection{LEO-LEO Channels} 
 For LEO we assume a pure vacuum is valid and that diffraction sets losses to 3\,dB for the 200\,km inter-separation distances and for  anticipated satellite transceiver apertures (of order 0.5m). However, we must bear in mind the satellites are moving rapidly in LEO and advanced tracking and routing control will be required. The GEO satellites will assist in these processes, largely via radio communications.
\subsubsection{GEO-LEO Channels} 
Although shown via two independent experiments that quantum communication is viable from GEO, both via quadrature measurements on coherent states and single-photon detection \cite{ref8}, here we consider that the large losses anticipated ($\sim60$\,dB) will render quantum communication from GEO of low likelihood. As discussed, we will consider only classical communication from GEO via radio links, noting GEO satellites can still play a critical role in the operation of a GQI.

A more quantitative assessment of  the LEO-Terrestrial and GEO-LEO links are given in Fig.~\ref{fig:rates} and Fig.~\ref{fig:rate2s}, respectively. These figures illustrate the best outcome for the entanglement distillation rate for both links, clearly demonstrating the impact of the much larger distance in any GEO-LEO link. This largely motivates the use of the latter links for classical communications only. However, it does remain possible that small quantities of quantum information transfer can be accommodated in the future via these long-distance links.

\subsection{Network Operations and Role of GEO Management Nodes}

\begin{figure}
	\centering
	\includegraphics[width=.95\linewidth]{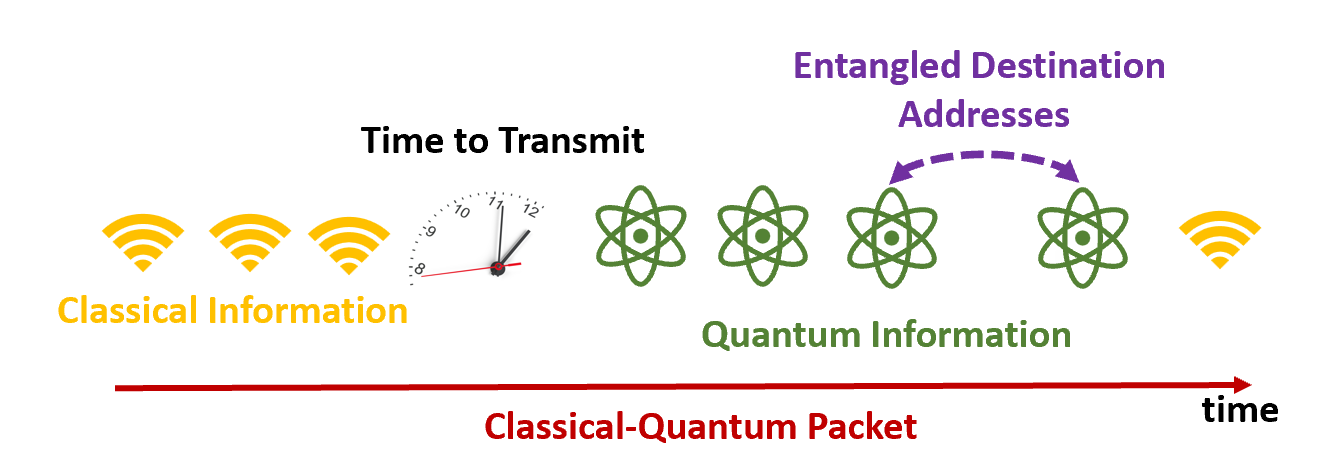}\label{fig:signal}
	\caption{A schematic of how the classical and quantum information can be formed into a packet. Here the quantum information could contain entangled qubits as a means of delivering superposition of paths for the quantum information transfer. {Note, the classical data and quantum data could originate from different photon sources, or different wavelengths, and multiplexed into a hybrid data frame at higher layers. The packet could also be constructed directly from a  photon source where some photons are transmitted in qubit form. Alternatively, the quantum data could simply be referenced as a memory address on the receiver. In fact, there are many ways to realize such a hybrid classical-quantum data structure for packet switching purposes in a future network.}	}
\end{figure}

{It is anticipated that  future quantum networks will follow a management structure based on the classical Internet - most notably an underlying communication architecture based on packet switching in which classical and quantum data are combined into single packets, e.g., \cite{packet}.}
Packet transfer will take many forms in our proposed architecture. Most novel in this respect will be the nature of the packets containing quantum information. Fig.~\ref{fig:signal} represents a potential form of the packets containing quantum information. This could be, for example, a packet being sent from a LEO satellite to a ground station as part of the initial entanglement distribution (Step 3 of Fig.~\ref{fig:network}). The classical information at the start of the packet header in this instance would contain, amongst other things, the requesting station ID, the receiving station ID,  the time of transmit, and/or the time to commence some quantum operation. The quantum information encapsulated in the packet could contain various forms of qubits. {Most importantly, some of these qubits could be entangled with each other, or even entangled with qubits in other packets - opening up the possibility passing teleportation-required resources through the network.
Entanglement in a general sense  would be a requirement for other quantum-only network applications such as  superposition of time ordering of path selections \cite{ref10}, or superposition of signal-paths  - commonly referred to as quantum routing \cite{routing}. 
Quantum routing is anticipated to be a key enabler of many important quantum applications in a future GQI - see \cite{routing} and references therein. As such, we view the ability to encapsulate entanglement information as a key aspect of future packet switching.} 
The final classical information appended at the end of the packet shown in Fig.~\ref{fig:signal} could contain error correction information and potential acknowledgement information, in addition to signalling the end of the quantum information. 

Different from classical networks, we should be aware that the different forms of information contained in quantum-network packets could be sent via different technologies and different frequencies (\textit{e.g.} radio and optical). The simplest way of allowing for this would be to have a highly synchronized network, coupled to very accurate positioning information (to allow accurate time of flight information to be obtained). Therefore, to enhance  current endeavours to enhance network synchronization via a variety of quantum techniques will be very important going forward \cite{ref11}.

For other network operations, packets will have similar structures to that shown in Fig.~\ref{fig:signal}. However, it is clear that some will forgo the need for quantum information encapsulation - such packets for  Step 1 of Fig.~\ref{fig:network}. These will largely contain the addresses of the required terrestrial stations that are to be involved in subsequent quantum information transfer.

Role of GEO nodes in our architecture is primarily for control. These nodes will contain the updated ephemeris of the LEO satellites, communicated directly to them by the LEO nodes, as well as responding to requests  from the terrestrial stations. Their principal role will be to dictate which LEO satellites will participate in the direct quantum entanglement distribution. Knowing the position information of all stations and satellites, the GEO node  will in most instances select the LEO satellites at the highest elevations (as seen from the relevant terrestrial stations). Position information will be pivotal in our architecture, not only for control and selection purposes, but also for application use in emerging quantum applications such as quantum enhanced GPS  and quantum location verification  \cite{ref12}. Although primarily used as control nodes via classical communication, it is possible that utilizing narrow-beam optical links some quantum information can be transferred between GEO and LEO satellites - albeit at much reduced transfer rates (see earlier discussions). 

Early implementations of CGI will use 
direct communication between terrestrial stations and LEO satellites - much akin to the operation of the quantum-enabled satellite Micius \cite{ref3}. However, going forward as the GQI grows in complexity and size (number of nodes)  the use of GEO satellites as outlined here can become standard. {Scalability will be an issue for all architectures as the network grows in size, but we suggest this will  not be a major issue in the first few  generations of the GQI. We point out, multiple GEO satellites will eventually be  deployed as part of the network (including GEO-GEO links), removing any centralization of the control plane.}

Integration of the space-based quantum communication links with terrestrial networks will largely be through teleportation between the terrestrial stations using the entanglement distributed via LEO. {We believe any optical fiber links between these stations will be limited in range, and possibly connected via quantum repeaters. Quantum repeaters are  devices that via entanglement swapping can increase the range of quantum links. These repeaters may also be used in inter-satellite links in the future, but we anticipate the first generations of GQI to only use direct entanglement distribution, and as such do not include discussion of them here.}

\begin{figure}
	\centering
	\includegraphics[width=.95\linewidth]{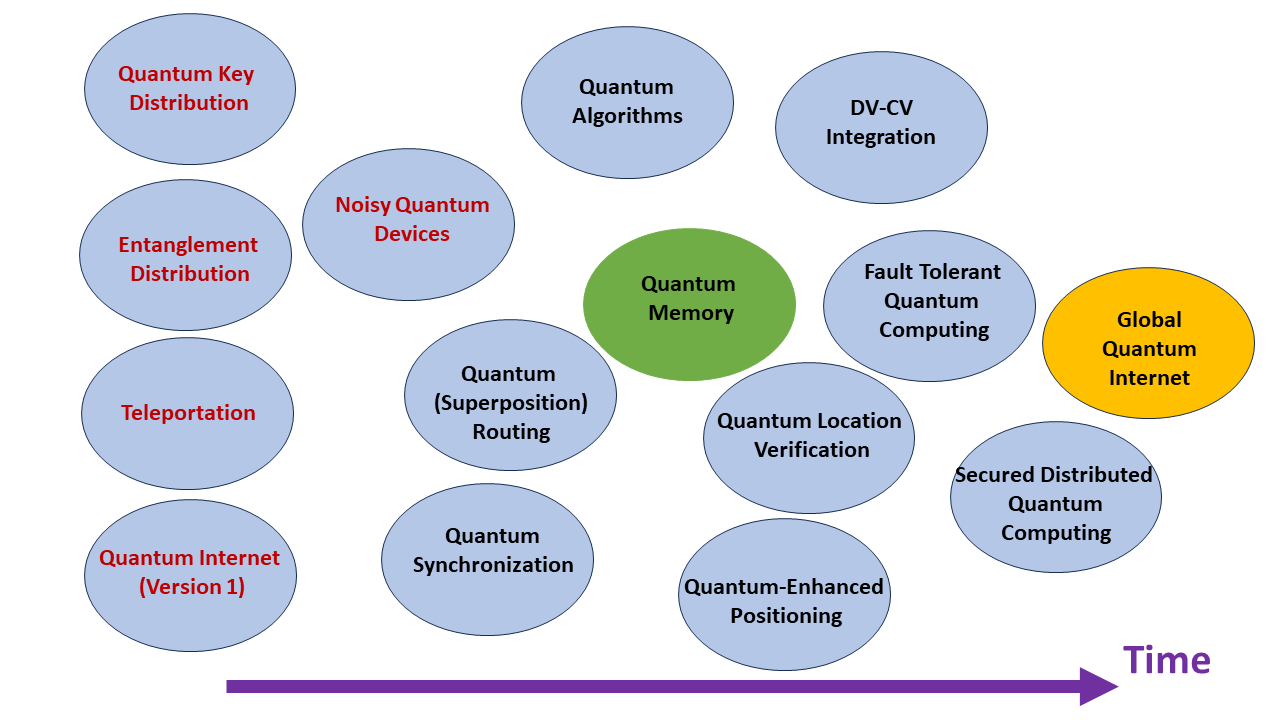}
	\caption{\textbf{The Evolution of the GQI.} Shown here is the estimated timeline to the full-blown GQI. The labels in red indicate the near term (current to 5 years) technical outcomes and applications. The green bubble indicates the importance of quantum memory in shaping the future Internet. The ultimate goal is shown in gold (10-20 years). {The `Quantum Internet [Version~1]' simply refers to technology already implemented by the Micius quantum-enabled satellite.}}
	\label{fig:apps}
\end{figure}

Our thoughts on the roll-out of the GQI are encapsulated in Fig.~\ref{fig:apps}. Here, in red we outline what we believe is in effect already available to us - and how these outcomes provide for the roll-out of a first-generation quantum Internet already.
We provide an overview of the future timeline towards the GQI. Pivotal for the ultimate goal of a full-blown quantum Internet will be the development of quantum memory - viable at room temperature and for a timescale of order round-trip times. Also, of course, full blown fault-tolerant quantum computers are required (the backbone of the GQI).  The good news is that exciting experimental progress is being made on both these fronts.

\subsection{Other Issues, Future Missions, and Research}\label{sec:QSD}
\newcommand{\qstate}{\M{\Xi}}
\newcommand{\qstateRV}{\RM{\Xi}}
\newcommand{\qmeas}{\M{\Pi}}
\newcommand{\rhoRV}{\qstateRV}
\renewcommand{\rho}{\qstate}

{A GQI will of course be inter-connected with terrestrial networks, and much can be learned from large-scale ground-based quantum networks already deployed \textit{e.g.} \cite{euro}. These networks largely deploy the killer-app of quantum communication - namely QKD. The advantages of this  application to security in next generation networks cannot be overstated as eventually this will be the security method of choice over all networked systems \cite{euro}. Many analyses on the economic benefits provided by the unconditional security offered by global networks underpinned by QKD, even considering the costs of building and maintaining such a quantum network, are given elsewhere \cite{costs}.
In addition, there are many quantum-based space mission currently anticipated to be launched in the next few years -see \cite{costs}. Although these will all be individual point-to-point missions, the interconnection of several quantum-enabled satellite are anticipated to be just a few years later. The number of launches provide some evidence that the timelines we propose in Fig.~\ref{fig:apps} are achievable.  Of special importance in this regard to future developments are the important elements of signal correction implementations in such networks such as distillation and quantum error correction. As these important quantum engineering issues are reviewed elsewhere \cite{errors}, we simply point out that experimental success on the ground should be readily transferred to space - albeit after serious consideration and design effort is give to SWAP (size, weight and power) issues. Currently, most of the new space missions will deploy DV quantum technology, with only a few dedicated to CV technology. The pros and cons of these competing technologies are detailed elsewhere \cite{DVCV}, but one of the main differences between these technologies in the space context relates to the methods utilized to combat channel conditions - i.e.  loss. In DV, this is quite simple, loss means no detection and the missed photon is simply ignored, whereas in CV the vacuum contribution enters to the quantum state at the receiver and that must be accounted for - a process that normally requires a detailed understanding of the turbulent channel conditions in the atmosphere - see \cite{costs} and references therein.  Our own belief is that eventually both technologies will co-exist in a hybrid network configuration.}

{Thus far we have avoided any detailed technical discussions on the various aspects of the GQI - this may lead to a false impression on the reader's part as to the complexity of such a network, and the amount of detailed ongoing research that  is being carried out in regard to various aspects of the GQI. The reality is the GQI will involve the deployment many complicated quantum tasks. In  an article such as this, it is impossible to discuss all such tasks. However, it is perhaps worthwhile to point to at least one of these tasks so as to provide some insight into the complexities that will underpin the GQI - and the nature of related ongoing quantum technology research. To this end, we choose \Ac{QSD}  - a task that
 aims to identify an unknown state among a set of quantum states \cite{moe6}, and is an essential component (building block) for several quantum applications including quantum communications, 
sensing, 
illumination, 
cryptography, 
control, 
and computing. \Ac{QSD} is of particular importance in the CV realm.}

\section{Conclusion}\label{sec:conclude}
The global quantum Internet will provide a large leap forward in communications for all. Pivotal to the  roll-out of this exciting new network will be the interaction of classical and quantum communications applied through an overlay of both terrestrial and space-based systems. In this article we have outlined some key issues regarding the architecture and performance metrics for this complex but exciting new communication system. We anticipate a first generation quantum Internet to be viable on a timescale of about 5 years and a full-blown quantum Internet in possibly two decades.

\bibliographystyle{IEEEtran}
\bibliography{mybib}

\end{document}

%% file: WINS-LatexInclusion/Acronyms/WINS-Acronyms.tex
\acrodef{BPC}[BPC]{broadening participation in computing}
\acrodef{BPE}[BPE]{broadening participation in engineering}
\acrodef{CISE}[CISE]{Computer and Information Science and Engineering}

\acrodef{PI}[PI]{Principal Investigator}
\acrodef{KI}[KI]{Key Investigator}
\acrodef{RO}[RO]{research objective}
\acrodef{RA}[RA]{research activity}
\acrodefplural{RA}[RA]{research activities}

\acrodef{STEM}[STEM]{science, technology, engineering, and mathematics}

\acrodef{FY}[FY]{fiscal year}
\acrodef{KPI}[KPI]{key performance indicator}

\acrodef{AFOSR}[AFOSR]{Air Force Office of Scientific Research}
\acrodef{ARO}[ARO]{Army Research Office}
\acrodef{DARPA}[DARPA]{Defense Advanced Research Projects Agency}
\acrodef{DOD}[DoD]{Department of Defense}
\acrodef{ONR}[ONR]{Office of Naval Research}
\acrodef{ISR}[ISR]{intelligence, surveillance and reconnaissance}
\acrodef{SA}[SA]{situational awareness}
\acrodef{MDMP}[MDMP]{military decision making process}
\acrodef{COA}[COA]{course of action}
\acrodef{OODA}[OODA]{observe, orient, decide, and act}

\acrodef{AA}[AeroAstro]{Aeronautics and Astronautics}
\acrodef{EECS}[EECS]{Electrical Engineering \& Computer Science}

\acrodef{SCC}[SCC]{Schwarzman College of Computing}
\acrodef{MIT}[MIT]{Massachusetts Institute of Technology}
\acrodef{IDSS}[IDSS]{Institute for Data, Systems, and Society}
\acrodef{ISN}[ISN]{Institute for Soldier Nanotechnologies}
\acrodef{LIDS}[LIDS]{Laboratory for Information and Decisions Systems}
\acrodef{WINSLAB}[WINS\,Lab]{Wireless Information and Network Sciences Laboratory}

\acrodef{OMEFAC}[OMEFAC]{Office of Minority Education's Faculty Advisory Committee}
\acrodef{MAES}[MAES]{Mexican American Engineers and Scientists}
\acrodef{HSSP}[HSSP]{high school studies program}

\acrodef{UROP}[UROP]{Undergraduate Research Opportunities Program}
\acrodef{MSRP}[MSRP]{MIT Summer Research Program}
\acrodef{URM}[URM]{underrepresented and minority}

\acrodef{IIT}[IIT]{Indian Institute of Technology}
\acrodef{SIT}[SIT]{Stevens Institute of Technology}
\acrodef{UA}[UA]{University of Arizona}

\acrodef{LIDAR}[LIDAR]{light detection and ranging}
\acrodef{NSTC}[NSTC]{National Science \& Technology Council}
\acrodef{NLA}[NLA]{noiseless linear amplification}
\acrodef{QIS}[QIS]{quantum information science}
\acrodef{QN}[QN]{quantum network}
\acrodef{QLS}[QLS]{quantum localization and synchronization}
\acrodef{QLSN}[{\it q}LSN]{quantum localization and synchronization network}
\acrodef{QNS}[QNS]{quantum network science}
\acrodef{QSD}[QSD]{quantum state discrimination}
\acrodef{RADAR}[RADAR]{radio detection and ranging}

\acrodef{AC}[AC]{actor-critic}
\acrodef{ACK}[ACK]{acknowledge}
\acrodef{AE}[AE]{angle estimate}
\acrodef{AI}[AI]{angle information}
\acrodef{AII}[AII]{angle information intensity}
\acrodef{AIS}[AIS]{automatic identification system}
\acrodef{AL}[AL]{angle likelihood}
\acrodef{ALU}[ALU]{arithmetic logic unit}
\acrodef{ANN}[ANN]{artificial neural network}
\acrodef{AOA}[AOA]{angle-of-arrival}
\acrodef{API}[API]{application programming interface}
\acrodef{ASIC}[ASIC]{application-specific integrated circuit}
\acrodef{AWGN}[AWGN]{additive white Gaussian noise}
\acrodef{AWS}[AWS]{Amazon Web Services}
\acrodef{BAC}[BAC]{Bayesian actor-critic}
\acrodef{BC}[BC]{belief condensation}
\acrodef{BCF}[BCF]{belief condensation filter}
\acrodef{BLE}[BLE]{Bluetooth Low Energy}
\acrodef{BP}[BP]{belief propagation}
\acrodef{BPZF}[BPZF]{band-pass zonal filter}
\acrodef{BTB}[BTB]{Bellini-Tartara bound}
\acrodef{BTZZB}[BTZZB]{Bellini-Tartara Ziv-Zakai bound}
\acrodef{CCDF}[CCDF]{complementary cumulative distribution function}
\acrodef{CDF}[CDF]{cumulative distribution function}
\acrodef{CE}[CE]{cross-entropy}
\acrodef{CEO}[CEO]{counting error outage}
\acrodef{CESS}[CE-SS]{cross-entropy SS}
\acrodef{CF}[CF]{characteristic function}
\acrodef{CFR}[CFR]{channel frequency response}
\acrodef{CIR}[CIR]{channel impulse response}
\acrodef{CNS}[CNS]{classical network science}
\acrodef{CR}[CR]{channel response}
\acrodef{CRB}[CRB]{Cram\'{e}r-Rao bound}
\acrodef{CRLB}[CRLB]{Cram\'{e}r-Rao lower bound}
\acrodef{CSD}[CSD]{Cauchy-Schwarz divergence}
\acrodef{CSI}[CSI]{channel state information}
\acrodef{CV}[CV]{continuous variable}
\acrodef{DA}[DA]{data-association}
\acrodef{DE}[DE]{distance estimate}
\acrodef{DFE}[DFE]{decision feedback equalizer}
\acrodef{DINS}[DINS]{decentralized inference in networked systems}
\acrodef{DNI}[DNI]{decentralized network inference}
\acrodef{DP}[DP]{dynamic program}
\acrodef{DPa}[DP]{direct path}
\acrodef{DPOMDP}[Dec-POMDP]{cecentralized-partially observed Markov decision process}
\acrodef{DRT}[DRT]{distance ratio test}
\acrodef{DV}[DV]{discrete variable}
\acrodef{EBIT}[EBIT]{entanglement bit}
\acrodef{ED}[ED]{energy detector}
\acrodef{EED}[EED]{Engineering Every Day}
\acrodef{EFIM}[EFIM]{equivalent Fisher information matrix}
\acrodef{EKF}[EKF]{extended Kalman filter}
\acrodef{EIRP}[EIRP]{equivalent isotropically radiated power}
\acrodef{ESD}[ESD]{energy-based soft-decision}
\acrodef{ESPRIT}[ESPRIT]{estimation of signal parameters via rotational invariant techniques}
\acrodef{FCC}[FCC]{Federal Communications Commission}
\acrodef{FG}[FG]{factor graph}
\acrodef{FII}[FII]{Fisher information inequality}
\acrodef{FIM}[FIM]{Fisher information matrix}
\acrodef{FL}[FL]{feature likelihood}
\acrodef{FP}[FP]{feature potential}
\acrodef{FSC}[FSC]{finite state controller}
\acrodef{FW}[FW]{Fisher-Wald}
\acrodef{FY}[FY]{fiscal year}
\acrodef{GDOP}[GDOP]{geometric dilution of precision}
\acrodef{GLMB}[GLMB]{generalized labeled multi-Bernoulli}
\acrodef{GLRT}[GLRT]{generalized likelihood ratio test}
\acrodef{GMMLMB}[G-MM-LMB]{Gaussian MM-LMB}
\acrodef{GNSS}[GNSS]{global navigation satellite system}
\acrodef{GP}[GP]{Gaussian process}
\acrodef{GPS}[GPS]{Global Positioning System}
\acrodef{HCA}[HCA]{heterogeneous computing architecture}
\acrodef{HDSA}[HDSA]{high-definition situation-aware}
\acrodef{HDP}[HDP]{hierarchical Dirichlet process}
\acrodef{HI}[HI]{hard information}
\acrodef{HMM}[HMM]{hidden Markov model}
\acrodef{IID}[IID]{independent, identically distributed}
\acrodef{IMU}[IMU]{inertial measurement unit}
\acrodef{INDFT}[INDFT]{inverse non-uniform discrete Fourier transform}
\acrodef{INR}[INR]{interference-to-noise ratio}
\acrodef{IOT}[IoT]{Internet-of-Things}
\acrodef{IR-UWB}[IR-UWB]{impulse radio UWB}
\acrodef{IRPR}[iRPR]{infinite regionalized policy representation}
\acrodef{JBSF}[JBSF]{jump back and search forward}
\acrodef{KDE}[KDE]{kernel density estimation}
\acrodef{KF}[KF]{Kalman filter}
\acrodef{KL}[KL]{Kullback-Leibler}
\acrodef{KLD}[KLD]{Kullback–Leibler divergence}
\acrodef{LBP}[LBP]{loopy belief propagation}
\acrodef{LEM}[LEM]{Laplacian eigen-map}
\acrodef{LEO}[LEO]{localization error outage}
\acrodef{LIDAR}[LIDAR]{light detection and ranging}
\acrodef{LMB}[LMB]{labeled multi-Bernoulli}
\acrodef{LMS}[LMS]{least means square}
\acrodef{LOCC}[LOCC]{local operations and classical communication}
\acrodef{LOS}[LOS]{line-of-sight}
\acrodef{LOT}[LoT]{Localization-of-Things}
\acrodef{LQG}[LQG]{linear-quadratic-Gaussian}
\acrodef{LRT}[LRT]{likelihood ratio test}
\acrodef{LRFS}[LRFS]{Labeled random finite set}
\acrodef{LS}[LS]{least squares}
\acrodef{LSE}[LSE]{line spectral estimation}
\acrodef{MAC}[MAC]{medium access control}
\acrodef{MAP}[MAP]{maximum a posteriori}
\acrodef{MBS}[MBS]{maximum bin search}
\acrodef{MC}[MC]{Monte Carlo}
\acrodef{MDD}[MDD]{minimum distance distribution}
\acrodef{MDP}[MDP]{Markov decision process}
\acrodef{MF}[MF]{matched filter}
\acrodef{MHT}[MHT]{multi-hypothesis tracking}
\acrodef{MIMO}[MIMO]{multiple-input multiple-output}
\acrodef{ML}[ML]{maximum likelihood}
\acrodef{MLE}[MLE]{maximum likelihood estimation}
\acrodef{MMSE}[MMSE]{minimum-mean-square-error}
\acrodef{MMLMB}[MM-LMB]{merged-measurement LMB}
\acrodef{MOT}[MOT]{multi-object tracking}
\acrodef{MOU}[MOU]{measurement origin uncertainty}
\acrodef{MP}[MP]{map potential}
\acrodef{MSE}[MSE]{mean-square error}
\acrodef{MUI}[MUI]{multi-user interference}
\acrodef{MUSIC}[MUSIC]{multiple signal classification}
\acrodef{NBI}[NBI]{narrowband interference}
\acrodef{NGF}[NGF]{NII generation function}
\acrodef{NIC}[NIC]{network inference encoding}
\acrodef{NISQ}[NISQ]{Noisy Intermediate-Scale Quantum}
\acrodef{NII}[NII]{network inference information}
\acrodef{NLN}[NLN]{network localization and navigation}
\acrodef{NLT}[NLT]{network localization and tracking}
\acrodef{NLOS}[NLOS]{non-line-of-sight}
\acrodef{NSF}[NSF]{National Science Foundation}
\acrodef{OOT}[OoT]{ocean-of-things}
\acrodef{OOTCESS}[OoT-CESS]{OoT-CESS}
\acrodef{OP}[OP]{outage probability}
\acrodef{OSPA}[OSPA]{optimum subpattern assignment}
\acrodef{P-Max}[P-Max]{$P$-Max}  
\acrodef{PAR}[PAR]{probabilistic association rule}
\acrodef{PCA}[PCA]{principal component analysis}
\acrodef{PDF}[PDF]{probability distribution function}
\acrodef{PDP}[PDP]{power delay profile}
\acrodef{PEB}[PEB]{position error bound}
\acrodef{PF}[PF]{physical features}
\acrodef{PHD}[PHD]{probability hypothesis density}
\acrodef{PMF}[PMF]{probability mass function}
\acrodef{POCS}[POCS]{projection onto convex sets}
\acrodef{POMDP}[POMDP]{partially observed Markov decision process}
\acrodef{PPM}[PPM]{pulse position modulation}
\acrodef{PPP}[PPP]{Poisson point process}
\acrodef{QUA}[QuaDRiGa]{QUAsi Deterministic RadIo channel GenerAtor}
\acrodef{RAT}[RAT]{radio access technology}
\acrodef{RCS}[RCS]{radar cross section}
\acrodef{RF}[RF]{radiofrequency}
\acrodef{RFID}[RFID]{radio frequency identification}
\acrodef{RFS}[RFS]{random finite set}
\acrodef{RI}[RI]{range information}
\acrodef{RII}[RII]{range information intensity}
\acrodef{RL}[RL]{reinforcement learning}
\acrodef{RLi}[RL]{range likelihood}
\acrodef{RLS}[R-LS]{range-based least squares}
\acrodef{RLMC}[RL-MC]{reinforcement learning Monte Carlo}
\acrodef{RM}[RM]{resource management}
\acrodef{RMS}[RMS]{root mean square}
\acrodef{RMSE}[RMSE]{root-mean-square error}
\acrodef{ROI}[ROI]{region of interest}
\acrodef{RPR}[RPR]{regionalized policy representation}
\acrodef{RRC}[RRC]{root raised cosine}
\acrodef{RSS}[RSS]{received signal strength}
\acrodef{RTT}[RTT]{round-trip time}
\acrodef{RV}[RV]{random variable}
\acrodef{SBS}[SBS]{serial backward search}
\acrodef{SBSMC}[SBSMC]{serial backward search for multiple clusters}
\acrodef{SC}[SC]{soft constraint}
\acrodef{SDE}[SDE]{stochastic differential equation}
\acrodef{SDN}[SDN]{software defined network}
\acrodef{SK}[SK]{soft knowledge}
\acrodef{SI}[SI]{soft information}
\acrodef{SII}[SII]{speed information intensity}
\acrodef{SIR}[SIR]{signal-to-interference ratio}
\acrodef{SLAM}[SLAM]{simultaneous localization and mapping}
\acrodef{SMC}[SMC]{sequential monte carlo}
\acrodef{SNR}[SNR]{signal-to-noise ratio}
\acrodef{SINR}[SINR]{signal-to-interference-plus-noise ratio}
\acrodef{SO}[SO]{soft observation}
\acrodef{SPAWN}[SPAWN]{sum-product algorithm over a wireless network}
\acrodef{SPEB}[SPEB]{squared position error bound}
\acrodef{SR}[SR]{sensor radar}
\acrodef{SS}[SS]{sensor selection}
\acrodef{SSCH}[SSh]{sensor scheduling} 
\acrodef{SVE}[SVE]{single-value estimate}
\acrodef{TBD}[TBD]{track before detect}
\acrodef{TCS}[TCS]{threshold crossing search}
\acrodef{TD}[TD]{temporal difference}
\acrodef{TDOA}[TDOA]{time difference-of-arrival}
\acrodef{TH}[TH]{time-hopping}
\acrodef{TNR}[TNR]{threshold-to-noise ratio}
\acrodef{TOA}[TOA]{time-of-arrival}
\acrodef{TOF}[TOF]{time-of-flight}
\acrodef{TSD}[TSD]{threshold-based soft-decision}
\acrodef{TWS}[TWS]{track while scan}
\acrodef{UAV}[UAV]{unmanned aerial vehicles}
\acrodef{UKF}[UKF]{unscendent Kalman filter}
\acrodef{UML}[UML]{unsupervised machine learning}
\acrodef{UWB}[UWB]{ultra-wideband}
\acrodef{VM}[VM]{von Mises}
\acrodef{VNA}[VNA]{vector network analyzer}
\acrodef{WAF}[WAF]{wall attenuation factor}
\acrodef{WBI}[WBI]{wideband interference}
\acrodef{WED}[WED]{wall extra delay}
\acrodef{WLS}[WLS]{weighted least squares}
\acrodef{WPAN}[WPAN]{wireless personal area network}
\acrodef{WSN}[WSN]{wireless sensor network}
\acrodef{WWB}[WWB]{Weiss-Weinstein bound}
\acrodef{XGL}[xGL]{next generation localization}
\acrodef{ZZB}[ZZB]{Ziv-Zakai bound}
\acrodef{ZZLB}[ZZLB]{Ziv-Zakai lower bound}

%% file: WINS-LatexInclusion/Definitions/WINS-Format-Def.tex
\definecolor{BLUE}{rgb}{0,0,1}

%% file: WINS-LatexInclusion/Definitions/WINS-Symbol-Def.tex
\DeclareMathAlphabet{\foo}{U}{tx-cal}{m}{n}

\newcommand{\hsa}[1] {h_{\text{s},\text{a}}}

\newcommand{\srxb}[1] {r_{\text{s-b},i}(t)}
\newcommand{\srxref}[1]{r_{\text{ref},i}(t)}
\newcommand{\srxrem}[1]{r_{\text{rem},i}(t)}

\newcommand{\thetaBi}[1]{\boldsymbol \theta_{B_i}}

\newcommand{\toar}[1]{\hat{\tau}_{i,0}}

\definecolor{myred}{rgb}{1,0.27,0}
\definecolor{mygreen}{rgb}{0.20, 0.8, 0.2}
\definecolor{myblue}{rgb}{0, 0, 0.8}
\definecolor{myorange}{RGB}{255, 178, 102}
\definecolor{mymagenta}{rgb}{0.78, 0.08, 0.52}
\definecolor{mycyan}{rgb}{0, 0.74, 1}

%% file: STQN-CM-V1.bbl
\begin{thebibliography}{10}
\providecommand{\url}[1]{#1}
\csname url@samestyle\endcsname
\providecommand{\newblock}{\relax}
\providecommand{\bibinfo}[2]{#2}
\providecommand{\BIBentrySTDinterwordspacing}{\spaceskip=0pt\relax}
\providecommand{\BIBentryALTinterwordstretchfactor}{4}
\providecommand{\BIBentryALTinterwordspacing}{\spaceskip=\fontdimen2\font plus
\BIBentryALTinterwordstretchfactor\fontdimen3\font minus \fontdimen4\font\relax}
\providecommand{\BIBforeignlanguage}[2]{{%
\expandafter\ifx\csname l@#1\endcsname\relax
\typeout{** WARNING: IEEEtran.bst: No hyphenation pattern has been}%
\typeout{** loaded for the language `#1'. Using the pattern for}%
\typeout{** the default language instead.}%
\else
\language=\csname l@#1\endcsname
\fi
#2}}
\providecommand{\BIBdecl}{\relax}
\BIBdecl

\bibitem{euro}
\BIBentryALTinterwordspacing
D.~Ribezzo, M.~Zahidy, I.~Vagniluca, N.~Biagi, S.~Francesconi, T.~Occhipinti, L.~K. Oxenlowe, M.~Loncaric, I.~Cvitic, M.~Stipcevic, Z.~Pusavec, R.~Kaltenbaek, A.~Ramsak, F.~Cesa, G.~Giorgetti, F.~Scazza, A.~Bassi, P.~De~Natale, F.~S. Cataliotti, M.~Inguscio, D.~Bacco, and A.~Zavatta, ``Deploying an inter-european quantum network,'' \emph{Advanced Quantum Technologies}, vol.~6, no.~2, p. 2200061, 2023. [Online]. Available: \url{https://onlinelibrary.wiley.com/doi/abs/10.1002/qute.202200061}
\BIBentrySTDinterwordspacing

\bibitem{ref2}
\BIBentryALTinterwordspacing
N.~Hosseinidehaj, Z.~Babar, R.~Malaney, S.~X. Ng, and L.~Hanzo, ``Satellite-based continuous-variable quantum communications: State-of-the-art and a predictive outlook,'' \emph{IEEE Communications Surveys \& Tutorials}, vol.~21, no.~1, pp. 881--919, 2019. [Online]. Available: \url{https://doi.org/10.1109/COMST.2018.2864557}
\BIBentrySTDinterwordspacing

\bibitem{ref3}
\BIBentryALTinterwordspacing
C.-Y. Lu, Y.~Cao, C.-Z. Peng, and J.-W. Pan, ``Micius quantum experiments in space,'' \emph{Rev. Mod. Phys.}, vol.~94, p. 035001, Jul 2022. [Online]. Available: \url{https://link.aps.org/doi/10.1103/RevModPhys.94.035001}
\BIBentrySTDinterwordspacing

\bibitem{costs}
\BIBentryALTinterwordspacing
L.~de~Forges~de Parny, O.~Alibart, J.~Debaud, S.~Gressani, A.~Lagarrigue, A.~Martin, A.~Metrat, M.~Schiavon, T.~Troisi, E.~Diamanti, P.~G{\'e}lard, E.~Kerstel, S.~Tanzilli, and M.~Van Den~Bossche, ``Satellite-based quantum information networks: use cases, architecture, and roadmap,'' \emph{Communications Physics}, vol.~6, no.~1, p.~12, Jan 2023. [Online]. Available: \url{https://doi.org/10.1038/s42005-022-01123-7}
\BIBentrySTDinterwordspacing

\bibitem{ref6}
\BIBentryALTinterwordspacing
E.~Villaseñor, M.~He, Z.~Wang, R.~Malaney, and M.~Z. Win, ``Enhanced uplink quantum communication with satellites via downlink channels,'' \emph{IEEE Transactions on Quantum Engineering}, vol.~2, pp. 1--18, 2021. [Online]. Available: \url{https://doi.org/10.1109/TQE.2021.3091709}
\BIBentrySTDinterwordspacing

\bibitem{ref5}
\BIBentryALTinterwordspacing
R.~Garc\'{\i}a-Patr\'on, S.~Pirandola, S.~Lloyd, and J.~H. Shapiro, ``Reverse coherent information,'' \emph{Phys. Rev. Lett.}, vol. 102, p. 210501, May 2009. [Online]. Available: \url{https://link.aps.org/doi/10.1103/PhysRevLett.102.210501}
\BIBentrySTDinterwordspacing

\bibitem{ref8}
\BIBentryALTinterwordspacing
K.~G\"{u}nthner, I.~Khan, D.~Elser, B.~Stiller, \"{O}mer Bayraktar, C.~R. M\"{u}ller, K.~Saucke, D.~Tr\"{o}ndle, F.~Heine, S.~Seel, P.~Greulich, H.~Zech, B.~G\"{u}tlich, S.~Philipp-May, C.~Marquardt, and G.~Leuchs, ``Quantum-limited measurements of optical signals from a geostationary satellite,'' \emph{Optica}, vol.~4, no.~6, pp. 611--616, Jun 2017. [Online]. Available: \url{https://opg.optica.org/optica/abstract.cfm?URI=optica-4-6-611}
\BIBentrySTDinterwordspacing

\bibitem{packet}
\BIBentryALTinterwordspacing
S.~DiAdamo, B.~Qi, G.~Miller, R.~Kompella, and A.~Shabani, ``Packet switching in quantum networks: A path to the quantum internet,'' \emph{Phys. Rev. Res.}, vol.~4, p. 043064, Oct 2022. [Online]. Available: \url{https://link.aps.org/doi/10.1103/PhysRevResearch.4.043064}
\BIBentrySTDinterwordspacing

\bibitem{ref10}
\BIBentryALTinterwordspacing
M.~Caleffi and A.~S. Cacciapuoti, ``Quantum switch for the quantum internet: Noiseless communications through noisy channels,'' \emph{IEEE Journal on Selected Areas in Communications}, vol.~38, no.~3, pp. 575--588, 2020. [Online]. Available: \url{https://doi.org/10.1109/JSAC.2020.2969035}
\BIBentrySTDinterwordspacing

\bibitem{routing}
\BIBentryALTinterwordspacing
A.~S. Cacciapuoti, J.~Illiano, and M.~Caleffi, ``Quantum internet addressing,'' \emph{IEEE Network}, pp. 1--1, 2023. [Online]. Available: \url{https://doi.org/10.1109/MNET.2023.3328393}
\BIBentrySTDinterwordspacing

\bibitem{ref11}
\BIBentryALTinterwordspacing
V.~Giovannetti, S.~Lloyd, and L.~Maccone, ``Quantum-enhanced positioning and clock synchronization,'' \emph{Nature}, vol. 412, no. 6845, pp. 417--419, Jul. 2001. [Online]. Available: \url{https://doi.org/10.1038/35086525}
\BIBentrySTDinterwordspacing

\bibitem{ref12}
\BIBentryALTinterwordspacing
R.~Malaney, ``The quantum car,'' \emph{IEEE Wireless Communications Letters}, vol.~5, no.~6, pp. 624--627, 2016. [Online]. Available: \url{https://doi.org/10.1109/LWC.2016.2607740}
\BIBentrySTDinterwordspacing

\bibitem{errors}
\BIBentryALTinterwordspacing
R.~Bala, S.~Asthana, and V.~Ravishankar, ``Combating errors in quantum communication: an integrated approach,'' \emph{Scientific Reports}, vol.~13, no.~1, p. 2979, Feb 2023. [Online]. Available: \url{https://doi.org/10.1038/s41598-023-30178-x}
\BIBentrySTDinterwordspacing

\bibitem{DVCV}
\BIBentryALTinterwordspacing
M.~Lasota, O.~Kovalenko, and V.~C. Usenko, ``Robustness of entanglement-based discrete- and continuous-variable quantum key distribution against channel noise,'' \emph{New Journal of Physics}, vol.~25, no.~12, p. 123003, dec 2023. [Online]. Available: \url{https://dx.doi.org/10.1088/1367-2630/ad0e8c}
\BIBentrySTDinterwordspacing

\bibitem{moe6}
\BIBentryALTinterwordspacing
S.~Guerrini, M.~Z. Win, and A.~Conti, ``Photon-varied quantum states: Unified characterization,'' \emph{Phys. Rev. A}, vol. 108, p. 022425, Aug 2023. [Online]. Available: \url{https://doi.org/10.1103/PhysRevA.108.022425}
\BIBentrySTDinterwordspacing

\end{thebibliography}
